\documentclass[runningheads]{llncs}
\usepackage[T1]{fontenc}

\usepackage{color,soul}
\usepackage{graphicx}
\usepackage{cite}
\usepackage{ulem}
\usepackage{amsmath}
\usepackage{amssymb}
\usepackage{subcaption} % for subfigures
\usepackage{geometry}
\usepackage{booktabs}
\usepackage{tabularx}
\usepackage{float}
\usepackage{bm}% Used for displaying a sample figure. If possible, figure files should

\begin{document}
\title{Multi-Agent Deep Q-Network with Layer-based 
Communication Channel for Autonomous Internal Logistics Vehicle Scheduling in Smart Manufacturing
}
\titlerunning{MADQN-LBCC for scheduling Internal Logistics}
\author{Mohammad Feizabadi\inst{1} \and
Arman Hosseini\inst{2}\and
Zakaria Yahouni\inst{1}}
\authorrunning{M. Feizabadi et al.}
\institute{Univ. Grenoble Alpes, CNRS, Grenoble INP G-SCOP, 38000 Grenoble, France \and
Systems Engineering, University of Virginia, Charlottesville VA 22903, USA
}
\maketitle              % typeset the header of the contribution
\begin{abstract}
{In smart manufacturing, scheduling autonomous internal logistic vehicles is crucial for optimizing operational efficiency. This paper proposes a multi-agent deep Q-network (MADQN) with a layer-based communication channel (LBCC) to address this challenge. The main goals are to minimize total job tardiness, reduce the number of tardy jobs, and lower vehicle energy consumption. The method is evaluated against nine well-known scheduling heuristics, demonstrating its effectiveness in handling dynamic job shop behaviors like job arrivals and workstation unavailabilities. The approach also proves scalable, maintaining performance across different layouts and larger problem instances, highlighting the robustness and adaptability of MADQN with LBCC in smart manufacturing.}
\keywords{Autonomous Internal Logistics, AIV scheduling, Dynamic flexible job shop, Multi-agent deep Q-network (MADQN), Layer-based communication channel (LBCC). }
\end{abstract}

\section{Introduction}

Internal Logistics Vehicles (ILVs) are crucial in enhancing the performance of manufacturing systems by facilitating the movement of products within manufacturing facilities \cite{soufi2024generation}. With the advent of Industry 4.0 technologies, the automation of these movements has been driven by multiple factors, including the improvement of production capacity and the reduction of injuries among human operators who traditionally moved heavy products. Notable technologies in this domain include Automated Guided Vehicles (AGVs) and Autonomous Intelligent Vehicles (AIVs), which autonomously transport products while considering workshop constraints and layouts.

The implementation of these systems presents several challenges that require careful consideration. Key among these challenges is the task of determining the priority for transporting products and selecting the appropriate vehicle for each transportation task. Moreover, these scheduling activities must account for various constraints, including delivery times, vehicle capacity and battery charging requirements, handling breakdowns of vehicles and machines, addressing urgent jobs, etc. Effective vehicle management is therefore crucial to align internal logistics with manufacturing objectives, such as minimizing tardiness of orders, achieving a balanced workload among workstations and vehicles, and optimizing the energy consumption of vehicles. To address these challenges, scheduling strategies consist of a set of rules designed to allocate vehicles to transportation requests while accounting for these complex constraints.

This paper addresses these challenges by introducing a multi-agent deep reinforcement learning approach along with a layer-based communication channel to dynamically allocate vehicles to transportation requests within a dynamically changing environment. Drawing from insights discussed in prior research by \cite{hosseini2024integrated}, which implemented a deep Q-network technique, this study makes several notable contributions. In the proposed multi-agent system, each job is instantiated as an individual agent, operating in a decentralized manner. These agents interact with one another to optimize their reward policies within a deep Q-network algorithm.  By leveraging this multi-agent framework, the system gains enhanced capabilities to navigate the complexities inherent in a dynamic manufacturing environment, characterized by the dynamic arrival of jobs and occurrences of workstation breakdowns/unavailibilites. 

The principal manufacturing objectives targeted by this approach encompass the minimization of total job tardiness, number of tardy jobs and the reduction of vehicle energy consumption. To assess the efficacy of our proposed methodology, we conducted a comparative analysis against various heuristic methods. Results from this evaluation demonstrate the superior performance of our multi-agent deep reinforcement learning method over traditional heuristics for the considered objectives. 

The paper is structured as follows: Section 2 delves into the literature, exploring related work regarding scheduling ILVs. Section 3 encompasses the problem description. Section 4 outlines the proposed multi-agent deep reinforcement learning approach. The subsequent sections are devoted to the experimental results, conducting analysis, and drawing conclusions.

\section{Related work}
 In this section, the state of the art in shop scheduling with intelligent transporters is first presented. Following this, the evolution and application of artificial intelligence and machine learning methods in these problems are reviewed. Finally, the current state of multi-agent systems applied to intelligent transporter scheduling in manufacturing shops is examined, with emphasis on the importance of communication between different agents and the integration of multi-agent systems with machine learning.
\subsection{Integrated shop and transporters scheduling}
Integrated shop and transporter scheduling is a comprehensive approach aimed at optimizing both the production schedule within a manufacturing shop and the logistics of material handling and transport. This approach considers the interdependencies between the resource assignment for production tasks to workstations and the transfer of jobs via vehicles such as AGV or AIV. 

In the study \cite{lee2005two}, the authors addressed two-machine flow shop and open shop scheduling problems where workpieces are transported via a single vehicle. For each problem, they proposed a heuristic algorithm aimed at minimizing the makespan. Notably, the transporter's capacity is assumed to be sufficient to transport any number of jobs, thereby neglecting the potential capacity constraints of vehicles in real-world scenarios.

More recent studies have increased the complexity of interstage transportation by considering multiple vehicles available in the field. This added complexity aims to better reflect real-world scenarios where multiple vehicles are used to transport workpieces between machines.
The study \cite{xue2018reinforcement},   tackled the multi-AGV flow shop scheduling problem using Q-learning to optimize makespan and minimize average job delays. However, this study does not consider the charging consumption of AGVs, which is a significant limitation in real-world scenarios.
In \cite{li2022multi}, authors explored the impact of AGV charging constraints on the scheduling of flexible manufacturing units with multiple AGVs. A genetic algorithm was employed to minimize the completion time while balancing AGV loads and reducing charging times. Their model effectively schedules AGVs to complete tasks and reduces charging wait times. However, this study overlooks the variations in job weights and their effects on transfer time and AGV charging consumption.

In real-world scenarios, another significant challenge complicating AGV scheduling is their load capacity, which is often overlooked in most studies. In \cite{li2023integrating}, authors considered transporter capacity limitations and layout rearrangement, proposing an improved non-dominated sorting genetic algorithm with hybrid local search (INSGA-HLS). Their approach successfully minimized exit time, labor cost, worker workload, and transportation time. 

%In \cite{shabtay2014combined} the authors addresse a bicriteria no wait flow shop scheduling involving multiple trnasporter robots. They suggest that the proposed case study can be reduced to a shortest path problem within a directed acyclic multigraph. Furthermore, 4 different variant of the problem with their solutions are proposed.  

\subsection{Artificial intelligence application in scheduling}
In the past few decades, there has been substantial growth in artificial intelligence (AI) applications within manufacturing, largely attributed to the accessibility of data facilitated by the Internet of Things (IoT). Supervised learning and reinforcement learning are among the most used methods in scheduling problems \cite{seeger2022literature}. Supervised learning relies on sample data, requiring data collection through real-world cases or simulations. A combination of supervised learning and linear programming was presented by \cite{jun2019learning}. A mixed-integer linear programming was initially employed to obtain the optimal solution, which served as the foundation for the supervised learning training. Subsequently, a random forest classifier was introduced to predict the priority of jobs. However, the study does not address the adaptability of the algorithm in dynamic environments with factors like machine breakdowns or stochastic processing times.
The authors \cite{hosseini2023scheduling} employed multiple linear regression to predict the optimal scheduling rule at each decision step in a dynamic job shop with a single AIV transporter. Their proposed method outperformed heuristic approaches in minimizing the makespan. It is important to note that supervised learning methods require substantial data collection and are not well-suited for real-time decision-making processes. 

Among the array of advancements in machine learning techniques, Deep Q-Networks stand out for their simplicity in implementation and their capacity to tackle complex problems by amalgamating deep learning with Q-learning \cite{hafiz2022survey}. Reinforcement learning (RL) is recognized as a broader framework encompassing decision-making tasks \cite{du2021survey}. A key distinction between RL and other AI algorithms lies in the learning mechanism, where RL learns through interactions with a dynamic environment, contrasting with supervised and unsupervised learning methods that rely on sample data for their learning process.

The authors \cite{zhao2021dynamic}, employed a Deep Q Network (DQN) to improve adaptive scheduling in dynamic job shops. The DQN algorithm selects actions from a set of ten heuristic dispatching rules. Their method demonstrated superior performance compared to single dispatching rules and traditional Q-learning. However, it is important to note that this work focused on a single performance indicator—total job tardiness—thereby overlooking the complexity of managing multiple objectives, which is more representative of real-world manufacturing environments.

The study \cite{zhou2022reinforcement}, utilized Deep Q-network (DQN) for online scheduling of a job shop, aiming to optimize multiple objectives including makespan, production cost, and machine utilization. Their approach effectively handled dynamic system behaviors such as urgent orders and machine failures and outperformed common scheduling methods such as Shortest Processing Time (SPT) heuristic and genetic algorithm. However, this study overlooked a significant complexity of real-world production systems: the optimization of job transportation. Modern manufacturing sites are equipped with intelligent vehicles, introducing new challenges for the production system.
In study \cite{hu2020deep}, the authors addressed transportation optimization on a flexible shop floor. They employed a Deep Q-Network (DQN) to schedule multiple AGVs in real-time, with the aim of minimizing the delay ratio and makespan. Two main decisions were handled: dispatching jobs and AGV selection. Their method outperformed previous approaches like AHP dynamic scheduling and traditional RL methods such as Q-learning and Sarsa. However, the study did not consider optimizing AGV charging consumption as an objective, nor did it address realistic limitations related to AGV charging, such as recharging times and policies, due to their added complexities.

In summary, the literature review highlights the significant advancements and applications of artificial intelligence, particularly in manufacturing scheduling. Several gaps are identified, including the optimization of multiple transportation vehicles with specific capacity constraints, considering their recharging times and policies, addressing multiple objectives, and accounting for unforeseen events such as machine breakdowns for robust optimization.

To adress these challenges, DQN offer a robust solution, merging deep learning with Q-learning to address complex scheduling challenges. Given its simplicity in implementation and proven effectiveness across various scheduling scenarios, our study adopts the DQN approach. The principles and workings of the Deep Q-Network are elaborated in the subsequent subsection.

\subsection{Deep Q-Network}

 Deep Q-Network (DQN) agent learns through experimenting within an environment\cite{arulkumaran2017deep}.  The base framework for such interaction is the Markov Decision Process which is represented by the tuple $(S,A,R)$ where:\\\\
$S$: Stands for the state space, $s_{t}\in S$ represents the state at time $t$\\
$A$: Stands for the action space, $a_{t} \in A$ represents action taken at time t considering $s_{t}$\\
$R$: stands for the reward function, $r_{t} \in R(s_{t},a_{t})$ represents the reward received by the agent after taking action $a_{t}$ in the state $s_{t}$\\\\
At each time step \(t\), the agent takes the action \(a_{t}\) and transitions from state \(s_{t}\) to state \(s_{t+1}\). The agent receives an immediate reward \(r_{t} = R(s_{t}, a_{t}, s_{t+1})\) upon reaching state \(s_{t+1}\) which indicates the performance of the agent. 
The objective of an RL algorithm is to maximize the cumulative rewards \cite{clifton2020q}. To achieve this objective, a DQN algorithm aims to approximate an optimal $ Q$ function of a given environment.
The optimal Q function $Q^*(s,a)$ represents the maximum expected cumulative reward that agent can obtain. Formally the Q function can be defined as :
\begin{equation}
Q(s, a) = \mathbb{E} \left[ \sum_{t=0}^\infty \gamma^t r_{t+1} \mid s_0 = s, a_0 = a \right]
\end{equation}

Where $\mathbb{E}$ denotes the expected value, $\gamma$ is the discount factor which lies in range [0,1], determining the importance of future rewards. And $r_{t+1}$ is the expected reward at the next time step. The optimal $Q$ function $Q^*(s,a)$ is formulated as:

\begin{equation}
    Q^*(s, a) = \max_\pi Q^\pi(s, a)
\end{equation}

To approximate $Q^*(s,a)$, DQN uses a deep neural network parametrized by $\theta$, which is denoted by $Q^{\theta} (s,a)$. The network takes the state $s_{t}$ as input and outputs the $Q$ values for all possible actions in the space.
The Q function is approximated using the temporal difference (TD) loss function \cite{hao2023exploration}. The main idea is to minimize the difference between the predicted Q values and the target Q values, which are computed using the following equation:
\begin{equation}
    Loss_{DQN}(\theta) = \mathbb{E}_{(s_{t},a_{t},r_{t},s^{'}_{t+1})} [y - Q^{\theta^{-}}(s_{t},a_{t})]^{2}
\end{equation}    \label{eq:TD}

%\[Q(s_t, a_t; \theta) \leftarrow Q(s_t, a_t; \theta) + \alpha \left( r_t + \gamma \max_{a'} Q(s_{t+1}, a'; \theta^-) - Q(s_t, a_t; \theta) \right),
%\]

Where $y=r_{t} + \gamma * max_{a^{'}}(Q(s^{'}_{t+1} , a^{'} ; \theta^{-}))$is the target value , $\gamma$  is the discount factor and  and $\theta^{-}$ represents the parameters of the target network, which are copied periodically from the primary network to stabilize training.

\subsection{Multi agent systems}
The review study conducted by \cite{bahrpeyma2022review} investigated the use of multi-agent reinforcement learning in smart factories, suggesting that multi-agent systems, when combined with AI techniques are the most appropriate for such environments. Furthermore, \cite{feriani2021single} provides a general background on single-agent and multi-agent deep reinforcement learning, highlighting the challenges and advantages of multi-agent systems over single-agent ones, particularly in wireless communications. 

One of the prominent challenges of Multi-Agent Deep Reinforcement Learning is the non-stationarity of the environment, which refers to the consistent variations in other agents' parameters. In the study \cite{ao2023energy}, authors targeted communication efficiency in multi-UAV cooperative trajectory planning. They applied a Double Stream Attention Multi-Agent Actor-Critic (DSAAC) algorithm. The study needs to enhance the algorithm’s adaptability to a broader range of dynamic scenarios and complex operational conditions.

In the study \cite{zhou2021multi}, a Multi-Agent Reinforcement Learning algorithm, specifically DQN, was employed to optimize operations. The agents within this system leverage Internet of Things (IoT) technology to facilitate inter-agent communication. The primary objectives of the study are to minimize the makespan and balance the total workload across the system. The results indicate that the distributed architecture inherent to the proposed approach effectively manages the high dimensionality characteristic of smart factory environments, outperforming traditional centralized methods. 

The authors \cite{maoudj2023capacitated} introduced a novel decentralized multi-agent system approach incorporating capacity constraints for AGVs. The agents collaborate in a decentralized manner to assign product transport tasks to each other and determine their routes. The results demonstrate that te proposed approach yields competitive outcomes compared to the Mixed-Integer Linear Programming model, highlighting its potential as a promising decentralized approach. However, the study primarily focuses on capacity constraints related to the weight of products, rather than the effects of transporting multiple products with a single AGV. 
 In the study \cite{wang2022solving}, the authors address a job scheduling problem in a resource preemption environment using a policy-based Multi-Agent Reinforcement Learning  approach. In this method, each job is considered an intelligent agent, utilizing a separate policy network. Additionally, a shared mixed Q-network is proposed to compute the global loss, fostering cooperation among the agents. Nevertheless, This study need to investigate the integration of diverse multi-agent scheduling  algorithms to tackle a wider range of Job Shop Scheduling Problems (JSSPs) and to enhance model generalization through advanced representation learning techniques.
 %It is worth noting that this study focuses solely on one manufacturing objective: makespan.

In summary, the literature review underscores the significant advancements and applications of AI in manufacturing scheduling, with a particular focus on the efficacy of multi-agent techniques and Deep Q-Networks in addressing complex challenges. Given the proven benefits of DQNs, our approach leverages multi-agent DQN to optimize Internal Logistics Vehicles scheduling.

\section{Problem description}
\label{sec:problem description}

For illustrative purposes, a case study is presented of a manufacturing system consisting of four assembly workstations (WS1, WS2, WS3, and WS4), a storage location, two charging stations, and two AIVs. These vehicles are utilized to transport raw materials from storage to workstations and to facilitate the movement of subassemblies and products between workstations. 

Four types of products (P1, P2, P3, and P4) are considered, each requiring processing at specific set of workstations. Each product undergoing processing at a workstation is referred to as an operation (the term job is also used for a product, which consists of a set of operations). The manufacturing shop floor is structured as a flexible job shop, where each product follows its own unique routing through the workstations. Additionally, the flexibility of the job shop allows certain operations to be handled by any one of several possible workstations rather than a single designated workstation. Table \ref{tab:routing} provides an overview of the different routings and workstation possibilities for each product operation.

\begin{table}

\centering

\caption{Routing of each product}
\label{tab:routing}

\begin{tabular}{|c|c|c|c|c|}
\hline
Products & Routs of Products \\
\hline
P1 &(WS1/WS3) -> (WS3/WS4) -> (WS2/WS3) \\
\hline
P2 & (WS1/WS4) -> (WS1/WS4) -> (WS2/WS3) \\
\hline
P3 & (WS4/WS5) -> (WS1/WS5) -> (WS2/WS3) \\
\hline
P4 & (WS1/WS4) -> (WS1/WS2) -> (WS3/WS4) \\
\hline
\end{tabular}
\end{table}

In real-world scenarios, manufacturing shops are not static; they must accommodate dynamic events such as the continuous arrival of jobs and the potential breakdowns of machines at workstations. These dynamic factors are considered in our problem.

\begin{figure}
    \centering
    \includegraphics[width=0.7\textwidth]{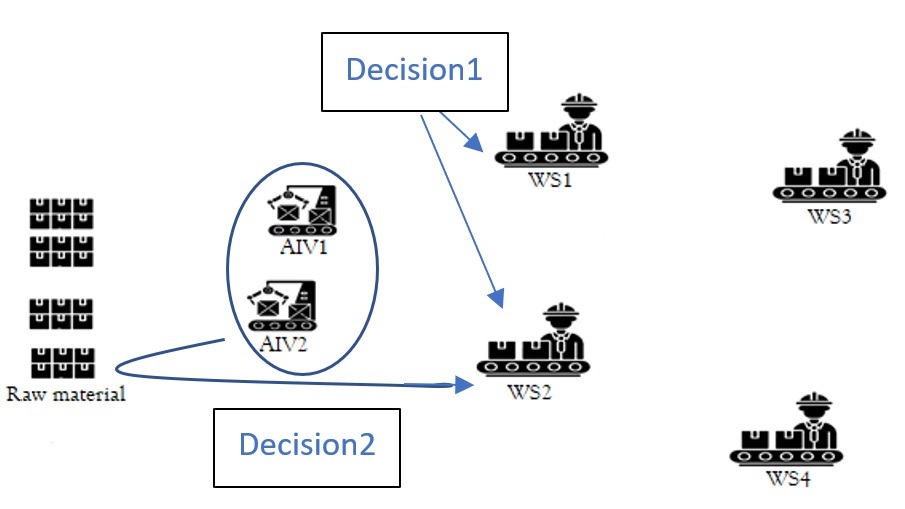}
    \caption{Illustration of the two decisions}
    \label{fig:Decisions}
\end{figure}

The goal of this study is to avoid tardiness to prevent penalty payments and optimize the use of AIVs to minimize energy consumption costs. As these orders arrive, each job (product) is scheduled individually, necessitating two key decisions as illustrated in Fig. \ref{fig:Decisions}:
\begin{enumerate}
     \item For each operation of a product, assign a workstation when multiple options are available. 
    \item For each product transfer to the designated workstation, assign an AIV 
%that is available and has the necessary capacity.  For the AIV assignment there is no constraint such as availablity or the remaining capacity of AIV
\end{enumerate}

%In a flexible job shop, an operation can be executed on various machines at different locations, each with distinct processing times. Consequently, the machine selection decision significantly impacts job tardiness and AIV energy consumption. Moreover, the presence of multiple AIVs with varying locations and charging levels further complicates the AIV selection process, making it a crucial factor in optimizing both objectives.

These two decisions must account for several constraints, including the remaining energy of each AIV, its loading capacity, the urgency of orders, the availability of AIVs, and the need to handle unexpected events such as the unavailability of workstations. 
Therefore, taking these two decisions ensures that products arrive at the workstations in a specific order and can be then processed. Transportation is influenced by the location and status of the AIVs. Consequently, these decisions directly impact the tardiness of jobs and the energy consumption of the AIVs. By optimizing these allocations, job tardiness can be effectively reduced, and energy usage minimized, thereby enhancing overall operational efficiency.

The job shop scheduling problem is known to be NP-hard \cite{yahouni2019evaluation}, and the additional complexity of allocating AIVs further exacerbates the problem. The challenge lies in determining the optimal allocation of AIVs and workstations such that the scheduling of jobs at workstations is optimal, all while managing the transportation and dynamic constraints. Consequently, advanced strategies are essential to effectively balance the multiple objectives and constraints inherent in this problem.

\section{Multi-agent deep Q-Network (MADQN) approach with layar-based communication channel (LBCC)}

In this section, a decentralized multi-agent Deep Q-Network (MADQN) approach is proposed to solve the illustrated problem described in section 3 involving AIV transporters. Furthermore, a layer-based communication channel (LBCC) is introduced and employed to deal with the non-stationarity of the multi-agent system. The primary objectives are to minimize total job tardiness, the number of tardy jobs,  and AIVs energy consumption. This approach addresses two types of decisions: \textit{workstation-selection} and \textit{AIV-selection}. Each job is considered an agent equipped with two deep Q-networks, one for each decision type.

All available jobs interact with the workshop environment simultaneously, resulting in the formation of queues for both AIVs and workstations. It is noteworthy that the queues for each AIV and workstation are managed based on a First-In-FirstOut (FIFO) policy, wherein jobs that are requested earlier are given priority.

\begin{figure}[ht]
    \centering
    \includegraphics[width=1.0\textwidth]{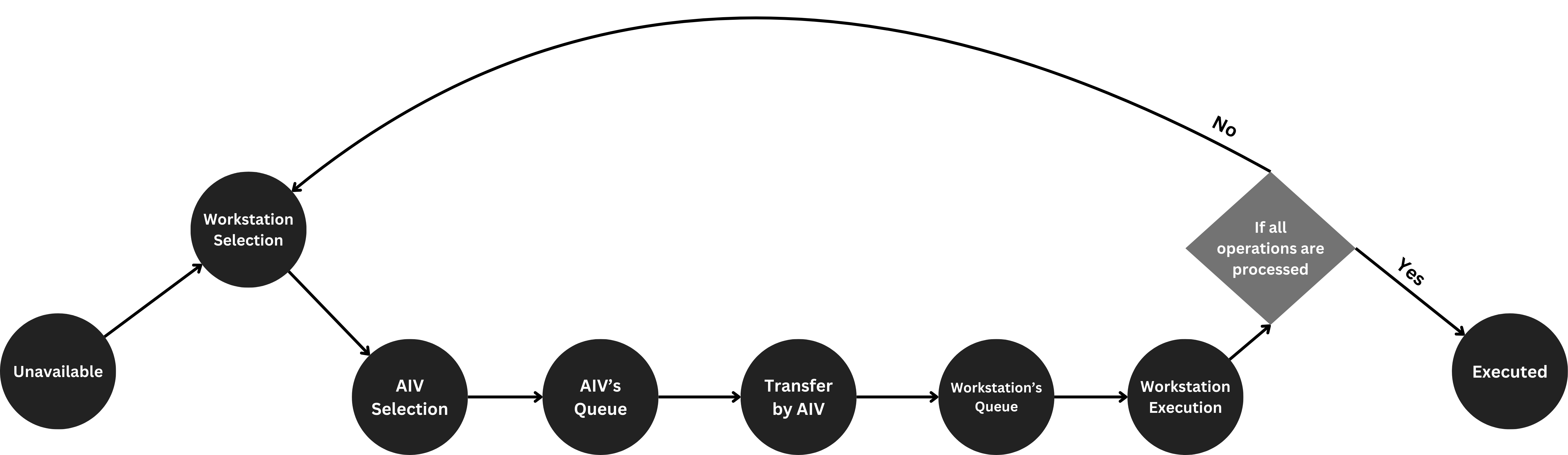}
    \caption{Graphical representation of the job's status}
    \label{fig:Status}
\end{figure}

For each job agent, two Deep Q-networks (DQN) are developed: one for \textit{workstation-selection} and the other for \textit{AIV-selection}. Each DQN is designed based on a Partially Observable Markov Decision Process. Status of each job are illustrated in Fig. \ref{fig:Status}.

The observations for each DQN are provided as follows:

\paragraph{\textbf{Workstation-selection DQN observation}}:
\begin{itemize}
    \item Queue length of workstations
    \item Distance of workstations from the job
    \item Workstations' busy time percentage
    \item Job's next processing time
    \item Job's current tardiness
    \item Job's remaining processing time
    \item Current time step
\end{itemize}
\paragraph{\textbf{AIV-selection DQN observation}}:
\begin{itemize}
    \item Queue length of AIVs
    \item Distance of AIVs from the job
    \item AIVs' battery percentage
    \item Job's current tardiness 
    \item Job's remaining processing time
    \item Current time step
\end{itemize}

%\paragraph{Actions}

\subsection{DQN characteristics: Actions, Rewards, Neaural Network features}
The action of the \textit{workstation-selection} DQN is to select the appropriate workstation for each available job. Similarly, the action of the \textit{AIV-selection} DQN is to select the appropriate AIV for transferring the job to the selected workstation.
It is crucial to manage the variation in the action space dimension for the \textit{workstation-selection} DQN at each step, which necessitates the use of a mask for the current operation. This mask delineates the feasible actions (workstations) available. For instance, if product/job P1 can only be processed on workstations WS2 or WS4 for its first operation among all $m$ workstations, other workstations are masked out except WP2 and WP4. Consequently, the network's output is adjusted to $-\infty$ where no viable action exists (masked options). This adjustment to  $-\infty$ is advantageous as it allows the objective function, which employs an argmax, to disregard the masked options.

%\paragraph{Reward}

At each step, two immediate rewards are allocated to the agent, one to the \textit{AIV-selection} DQN and another to the \textit{workstation-selection} DQN. The \textit{AIV-selection} DQN receives a reward immediately upon completing the transfer of a product to the designated workstation. This reward is computed based on the energy consumption of the AIV, encompassing the period from when the AIV commences the job pick-up process until it delivers the product. This reward is assigned a negative value. 
The \textit{workstation-selection} DQN obtains a reward immediately after completing the execution of the product's operation. This reward is calculated based on the current tardiness of the product upon finishing that operation, which is computed as follows for each product:

\begin{equation}
    Current Tardiness_{i} = max(0,k_{i}\times RPT_{i} + CT  - DueDate_{i}), 
    \quad i = 1, 2, \ldots, n;
\end{equation}

Where $RPT_i$ stands for the Remaining Processing Time of $product_i$, $CT$ is the current time of the simulation, and $k$ is a coefficient used to account for the transfer times of the job. This reward, given as a negative value, reflects the deviation from the due date, thereby incentivizing the minimization of product/job tardiness.

Moreover, a final reward is assigned to each job agent for both DQNs when the product is executed completely (i.e., all operations of the product are completed). The \textit{final reward} is calculated based on the lateness of each product after the product is executed. For each $product_i$ lateness is defined as:

\begin{equation}
    Lateness_{i} =   CompletionTime_{i} - DueDate_{i} \quad i = 1, 2, \ldots, n;
\end{equation}

\begin{equation}
    FinalReward_{i} = Lateness_{i} \times (-1) \quad i = 1, 2, \ldots, n;
\end{equation}

%\paragraph{Neural Networks}
The structure of all networks is identical. A fully connected deep neural network with five hidden layers is utilized, with each layer consisting of 10 nodes and employing the Tanh activation function. It is noteworthy that all inputs for the agents are normalized. Stochastic Gradient Descent is employed to optimize the Mean Squared Error loss function. The learning rates are uniformly set to 0.01, and the discount factors are set to 0.9.

\subsection{Layer Base Communication Channel (LBCC)}
A key advantage of multi-agent systems lies in their capacity for inter-agent communication, which substantially enhances system performance. In the proposed approach, the uniform structure of all agents facilitates the creation of a communication channel across corresponding layers. This communication channel is integrated into the networks of the job agents' DQNs (\textit{workstation-selection} and \textit{AIV-selection}). Specifically, within each hidden layer of a job agent's network, the input to that hidden layer, originating from the output of the preceding layer, is combined with the output of the same hidden layer from all other job agents. This design ensures that each job agent is aware of the outputs of other agents at the same level of the hidden layer. Such an integrated communication mechanism promotes effective information sharing and coordination among agents, enhancing overall system efficiency. The mathematical representation of the communication channel is provided below:

%\[h^{(l)}_{i} = \tanh\left(W^{(l)} \cdot \left(h^{(l-1)}_{i} +\ \sum_{j \neq i} h^{(l)}_{j}\right) + b^{(l)}\right)\]
\begin{equation}
    h^{(l)}_{out,i} = \tanh\left(W^{(l)} \cdot \left(h^{(l)}_{in,i} \cup\  H^{(l)}\ \right) + b^{(l)}\right) 
 \quad i = 1, 2, \ldots, n;
\end{equation}

Where:
\begin{itemize}
%\( h^{(l)}_{out,i} \) represents the output of the \( l \)-th hidden layer of the \( i \)-th job agent.

    \item \( h^{(l)}_{out,i} \) is the output of \( l \)-th layer of the \(i\)-th job agent.    
    \item \( W^{(l)} \) is the weight matrix for the \( l \)-th layer.
    \item \( h^{(l)}_{in,i} \) is the input to the \( l \)-th layer of the \( i \)-th agent.
    \item $H^{(l)}\ = \ \cup_{j \neq i}\ h_{j}^{(l)}$ represents the union of the outputs from the same hidden layer \( l \) of all other job agents \( j \) except \( i \).
    \item \( b^{(l)} \) is the bias vector for the \( l \)-th layer.
    
%\( \sum_{j \neq i} h^{(l)}_{j} \) represents the sum of the outputs from the same hidden layer \( l \) of all other job agents \( j \) except \( i \).
    \item \( \tanh \) is the Tanh activation function applied to the linear combination of the inputs, weights, and biases.
\end{itemize}

\begin{figure}[ht]
    \centering
    \includegraphics[width=0.8\textwidth]{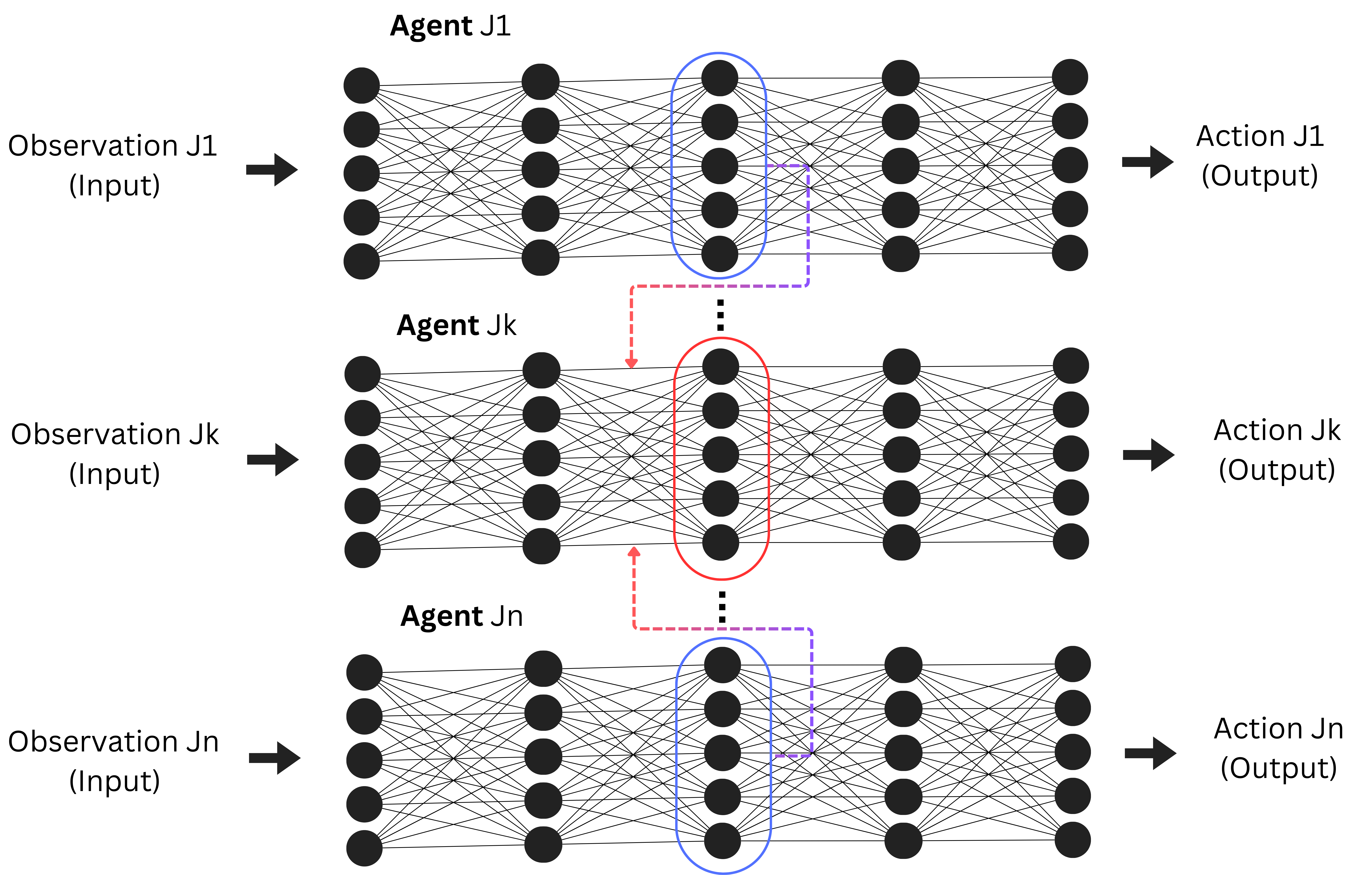}
    \caption{Communication channel example for the third layer of \textit{workstation-selection} network}
    \label{fig:Channel}
\end{figure}

As an example, Fig. \ref{fig:Channel} illustrates the communication mechanism within the \textit{workstation-selection} networks of job agents at the third hidden layer. Specifically, for job $J_k$, the input to its third hidden layer is integrated with the outputs from the third hidden layer of all other jobs, ranging from $J_1$ to $J_n$ and $1 < k < n$.

\section{Experimentation and results}
\subsection{Case study data}

The case study described in Section \ref{sec:problem description}, which involves four products and four workstations, is used to experiment the MADQN approach. In this example, each product can have various numbers of job arrivals (each job represent one product). It is assumed that the number of jobs for all product types is equal. For instance, in the case of a total of 60 job arrivals, each of the four product types has 15 job arrivals. The time between the arrival of jobs is assumed to follow an exponential distribution with a rate parameter of $\lambda = 5$ time units. 

%In dynamic job shop problems, job arrivals follow a Poisson distribution closely \cite{shahrabi2017reinforcement}, and the time between the arrival of jobs are assumed to be exponentially distributed 

Dealing with tardiness as two of the three  objectives requires having due dates for each job. In this case study, a due date generator is used based on the work of \cite{adibi2010multi} for each job. The due date of job $i$ is computed as shown in the following equation:
\begin{equation}
    DueDate_i = ArrivalTime_i + T \sum_{}^{ } MeanProcessingTime_{i}, \quad i = 1, 2, \ldots, n;
\end{equation}

Essentially, the due date is equal to the arrival time of the job plus the total required processing time for that job. T is a coefficient to take into account the transfer time of jobs between workstations. This coefficient follows a normal distribution with parameters defined as follows :

\[
\mu=\frac{Nb\_Jobs}{4}\quad,\quad \sigma=4
\]
As the problem is a flexible job shop, jobs have the flexibility to be processed on different workstations for their various operations. Therefore, the processing time for each operation of the job is determined by the mean of the processing times of that operation on the possible workstations. For example, if one operation can be processed on either workstation WS1 with a processing time of 8 or on workstation WS2 with a processing time of 4 minutes, $MeanProcssingTime$ for that operation will be 6 minutes.

To enhance the dynamic behavior of the job shop, machine breakdowns or in this case workstation unavailibilities are also considered. An exponential distribution is applied for both the time interval between two unavailabilities (TBI) and the time required for fixing the problem (workstation to be available) (TRF) \cite{shahrabi2017reinforcement}. The parameters for TBI (time between two inspections) and TRF (time required for fixing) are set to $\lambda_{\text{TIB}} = 200$ and $\lambda_{\text{TRF}} = 50$.

Another important aspect is the layout of the job shop, which is defined by the distances between workstations, the raw material storage location, and the charging stations. To generate different layouts, a uniform random variable between 10 and 50 is used for each transfer time, reflecting various possible distances and assuming that both AIVs have the same speed.

Each of the two AIVs is set with a loading capacity of two, and each job/product occupies one out of the two capacity units of the AIVs. In other words, AIVs are capable of transferring two products simultaneously. The transfer policy is that AIVs fill their capacity if there are enough requests; otherwise, they collect the available requests, carry all picked-up products to their destinations, and then become idle at the last destination, ready for the next transfer.

There are two charging stations each with a capacity of one AIV at the same time. AIVs are sent to the charging station when their battery level falls below 40\%. The location of the charging stations can vary across different layout setups. The charging consumption of the AIVs is assumed to correlate with their status (moving or stopped) and the number of product they are carrying. The energy consumption percentage pattern of AIVs is provided in Table \ref{tab:aiv_consumption}.

\begin{table}[ht]
    \centering
    \caption{AIV Consumption Based on Carrying Load}
    \begin{tabular}{cc}
        \toprule
        \textbf{Condition} & \textbf{Consumption} \\
        \midrule
        Not moving            & 0.01\%\\
        Moving, carrying 0 product & 0.02\% \\
        Moving, carrying 1 product  & 0.05\% \\
        Moving, carrying 2 products & 0.10\% \\
        \bottomrule
    \end{tabular}
    
    \label{tab:aiv_consumption}
\end{table}

\subsection{Results}

The proposed case study is unique and novel due to the added complexities, constraints, and dynamic events. Since there is no existing solution in the literature to compare our approach with, nine heuristics are provided to evaluate the performance of the proposed MADQN method. These heuristics are generated by mixing three dispatching rules for \textit{workstation-selection} and three rules for \textit{AIV-selection}. The three dispatching heuristics for \textit{workstation-selection} are as follows:

\begin{itemize}

    \item \textbf{SPT (Shortest Processing Time)}: Workstation with the Shortest Processing Time is selected for each product. It is worth noting that the processing time of each operation varies depending on the workstation performing the process.
    \item \textbf{SQL (Shortest Queue Length)} : Workstation with the shortest queue length has priority.
    \item \textbf{SWL$_W$ (Shortest Workload)}: Workstation with the Shortest Workload or shortest percentage of busy time has priority.

\end{itemize}

The three dispatching rules for \textit{AIV-selection} are:

\begin{itemize}
    \item \textbf{MC (Most Charge)}: Select the AIV with the most remaining charge.
    \item \textbf{STT (Shortest Transfer Time)}: Select the AIV with the Shortest Transfer Time (from its current location to the location of the first product to pick up).
    \item \textbf{SWL$_A$ (Shortest Workload)}: Choose the AIV with the Shortest Workload or shortest percentage of busy time.
\end{itemize}

%Two types of approaches are utilized to evaluate the effectiveness of the proposed method: scalability and generalization on four workstations-two AIVs setup. Scalability is typically assessed by evaluating the method across different ranges of arriving products. As the number of products increases, the complexity of the scheduling problem also rises, making it more challenging to optimize both objectives. A set of simulations with an increasing number of products (20, 40, 60, 80 and 100) is performed. For each product count, the simulation is executed 100 times (each time with different values of processing times and time arrivals of products), and the average performance result is calculated. The objectives of the problems are set to the total tardiness of all products and the total AIVs' charging consumption. Figure \ref{fig:Lateness} represents the total tardiness of MADQN compared to the nine combinations of heuristics with different numbers of products. It is evident that the superiority of MADQN is even more pronounced as the number of products increases. Additionally, the total tardiness of the MADQN method remains relatively stable and does not exhibit significant variation.

The effectiveness of the proposed approach is evaluated based on jobs scalability. Scalability is typically assessed by evaluating the method across different ranges of arriving products. As the number of products increases, the complexity of the scheduling problem also arises, making it more challenging to optimize the three objectives. A set of simulations with an increasing number of jobs (20, 40, 60, 80, and 100) is conducted. For each job count, a unique layout is considered, which is based on the transfer times between workstations and process times of operations on workstations. This layout remains consistent for each specific job count but varies between different job counts. The simulation is executed 100 times (each time with different values of processing times and time arrivals of products/jobs).

The objectives of the problems are set to the total tardiness of all products, the number of tardy products, and the total AIVs' charging consumption.
For each objective, the mean results of all 100 simulations are collected for each job count. Based on Table \ref{tab:tardiness}, \ref{tab:nb_tardy}, and \ref{tab:consumption}, where best results are highlighted in bold, the proposed MADQN method outperforms all nine heuristic combinations across all three objectives. Importantly, the results remain stable as the number of jobs increases, even when the job shop layouts change and different processing times are used. This stability demonstrates the scalability of the proposed approach.

\begin{table}[ht]
    \centering
    \caption{Mean of total tardiness for each 100 simulation of n jobs}
    \label{tab:tardiness}
    \tiny
    \begin{tabular}{c|c|c|c|c|c|c|c|c|c|c}
    \toprule
    Jobs & $STT.SPT$ & $STT.SQL$ & $STT.SWL_W$ & $SWL_A.SPT$ & $SWL_A.SQL$ & $SWL_A.SWL_W$ & $MC.SPT$ & $MC.SQL$ & $MC.SWL_W$ & $MADQN$ \\
    \midrule
     20 & 4084.7 & 3755.9 & 5382.5 & 4537.6 & 3626.2 & 5134.3 & 7385.0 & 4068.2 & 5148.8 & \textbf{2586.7}\\
    40 & 33520.4 & 30683.9 & 31601.8 & 21104.5 & 34829.9 & 30927.5 & 28467.1 & 37616.1 & 38026.5&\textbf{7917.2} \\
    60 & 46410.3 & 57723.2 & 74085.8 & 26591.7 & 13932.4 & 43301.0 & 32431.8 & 28510.7 & 57797.8 &\textbf{10229.2}\\
    80 & 96602.7 & 146211.9 & 198746.6 & 35938.4 & 58552.9 & 83505.9 & 41344.1 & 48923.0 & 75331.7& \textbf{14546.6} \\
    100 & 45686.28&110800.91&	134116.17	&32336.17&	80030.2&	80112.83&38905.59	&77558.05&	91913.11&\textbf{19363.87}\\

    \bottomrule
    \end{tabular}
\end{table}

\vspace{-2em}

\begin{table}[ht]
    \centering
    \caption{Mean of number of tardy jobs for each 100 simulation of n jobs}
    \label{tab:nb_tardy}
    \tiny
    \begin{tabular}{c|c|c|c|c|c|c|c|c|c|c}
    \toprule
    Jobs & $STT.SPT$ & $STT.SQL$ & $STT.SWL_W$ & $SWL_A.SPT$ & $SWL_A.SQL$ & $SWL_A.SWL_W$ & $MC.SPT$ & $MC.SQL$ & $MC.SWL_W$ & $MADQN$ \\
    \midrule
     20 & 14.76 & 14.32 & 16.20 & 16.53 & 14.45 & 17.49 & 17.95 & 15.63 & 17.01 & \textbf{12.40} \\
    40 & 32.99 & 37.91 & 38.44 & 34.52 & 38.20 & 37.28 & 34.28 & 36.81 & 36.46 & \textbf{19.32} \\
    60 & 42.75 & 42.54 & 47.77 & 42.41 & 30.94 & 48.54 & 42.64 & 36.98 & 48.46 & \textbf{24.39} \\
    80 & 58.10 & 71.99 & 74.80 & 47.61 & 58.32 & 64.85 & 47.10 & 50.78 & 56.51 & \textbf{36.74} \\
    100 & 62.56 & 70.96 & 74.64 & 57.4 & 74.13 & 74.16 & 56 & 66.15 & 67.67 & \textbf{43.61} \\
    \bottomrule
    \end{tabular}
\end{table}

\begin{table}[ht]   
    \centering
    \caption{Mean of total energy consumptions for AIVs for each 100 simulation of n jobs}
    \label{tab:consumption}
    \tiny
    \begin{tabular}{c|c|c|c|c|c|c|c|c|c|c}
    \toprule
    Jobs & $STT.SPT$ & $STT.SQL$ & $STT.SWL_W$ & $SWL_A.SPT$ & $SWL_A.SQL$ & $SWL_A.SWL_W$ & $MC.SPT$ & $MC.SQL$ & $MC.SWL_W$ & $MADQN$ \\
    \midrule
    20 & 95.77 & 92.37 & 102.12 & 98.15 & 91.19 & 105.33 & 103.71 & 88.80 & 101.64 & \textbf{76.86} \\
    40 & 235.24 & 262.89 & 261.62 & 236.13 & 283.86 & 270.66 & 241.24 & 270.53 & 273.47 & \textbf{166.76} \\
    60 & 320.55 & 341.63 & 393.10 & 320.58 & 281.60 & 370.85 & 319.18 & 291.34 & 392.81 & \textbf{240.97} \\
    80 & 444.40 & 457.23 & 562.18 & 395.93 & 444.85 & 508.96 & 399.52 & 417.69 & 476.92 & \textbf{323.53} \\
    100&410.69 & 508.54 & 537.94 & 401.46 & 498.74 & 511.77 & 403.18 & 488.30 & 518.54 & \textbf{348.04} \\
    \bottomrule
    \end{tabular}
\end{table}

\begin{figure}[ht]
    \centering
    \includegraphics[width=1.0\textwidth]{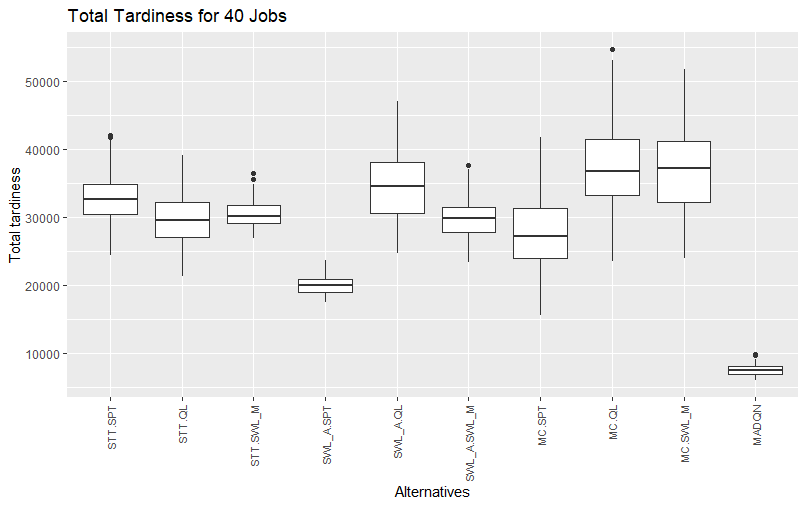}
    \caption{Total tardiness of 40 product counts}
    \label{fig:tardinessBox}
\end{figure}

\begin{figure}[ht]
    \centering
    \includegraphics[width=1.0\textwidth]{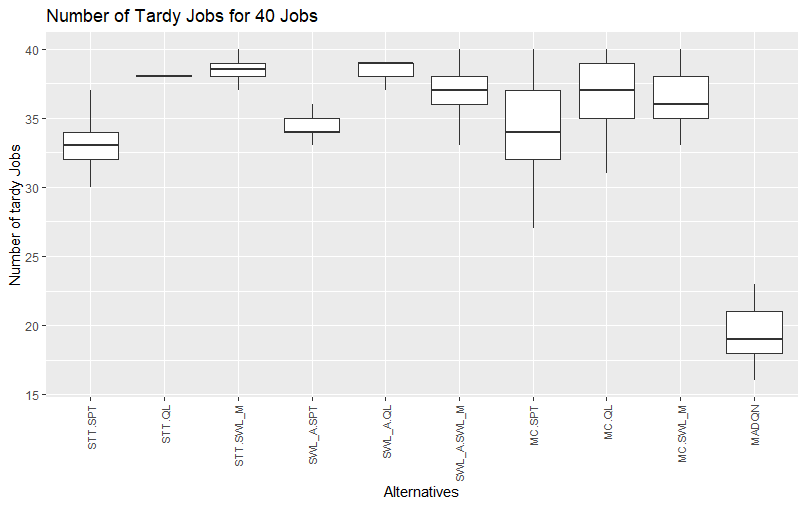}
    \caption{Number of tardy jobs of 40 product counts}
    \label{fig:nb_tardyBox}
\end{figure}

\begin{figure}[H]
    \centering
    \includegraphics[width=1.0\textwidth]{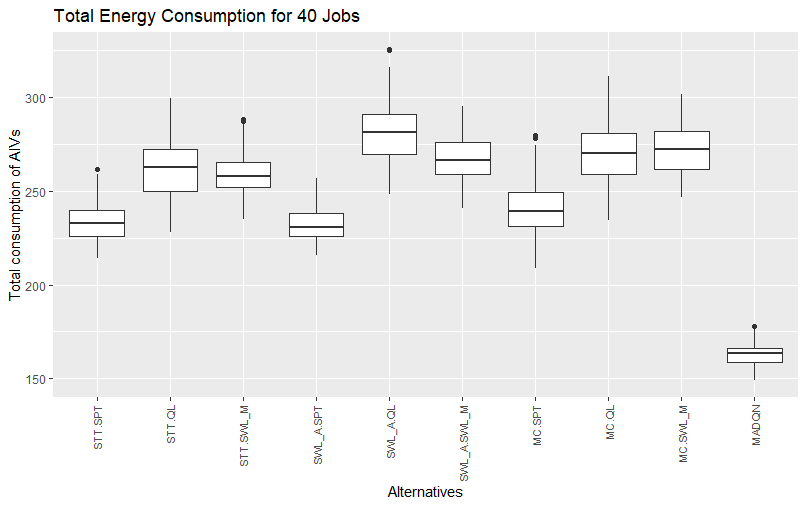}
    \caption{AIVs total energy consumption}
    \label{fig:consumptionBox}
\end{figure}

To provide a comprehensive comparison between the distribution of results obtained by MADQN and other heuristics, box plots for all three objectives are presented for a product count of 40. Due to space constraints, plots for other product counts are not included. Fig. \ref{fig:tardinessBox}, \ref{fig:nb_tardyBox}, and \ref{fig:consumptionBox} demonstrate the notable disparity in the distribution of results between MADQN and other heuristics. Regarding total tardiness, a comparison of the median values reveals that MADQN is centered around 5000 time-units, while the best-performing heuristic, $SWL_A.SPT$, is centered around 20000 time-units. This approximate 15000 time-unit gap in tardiness, can have a significant impact on manufacturing delay costs. Furthermore, the small variations in values for MADQN are reflected in the box plot, which shows that quartiles are close together. This minimal variation underscores the reliability and stability of the results across the 100 experiments.

For the second objective as shown in Fig. \ref{fig:nb_tardyBox}, the number of tardy jobs, MADQN values are clustered around 18, while the best performing heuristic in this regard, $STT.SPT$, centers around 33. With 40 jobs, employing MADQN enables the timely delivery of approximately 15 more jobs compared to the best heuristic, which can have a substantial impact on the costs associated with tardy jobs.

In the case of the third objective, total AIV energy consumption, MADQN results center around 163\% of battery percentage consumption, whereas heuristics, $STT.SPT$ and $SWL_A.SPT$, centers around 235\% (Fig. \ref{fig:consumptionBox}). This difference not only indicates energy savings but also translates into time savings within the job shop.
According to the recharging policy assumption, AIVs are sent to the charging station when their battery level falls below 40\%. In other words, each 60\% battery consumption necessitates a recharge. With the $SWL_A.SPT$ heuristic, approximately 4 recharging instances are required, while with MADQN, only around 2 recharges occur. Considering that each recharge takes approximately 30 time units and temporarily suspends transfers between workstations, the reduced number of recharges with MADQN contributes to lower job tardiness.

\subsection{Discussion}
The proposed approach effectively addresses several key challenges in real-world production systems, particularly in flexible job shop scheduling where multiple routing options and dynamic events, such as machine breakdowns, stochastic job arrivals, and variable processing times, complicate decision-making. The integration of Autonomous Intelligent Vehicles (AIVs) for material handling, while enhancing efficiency, introduces constraints like energy management and battery recharging. Notably, our study incorporates the often-overlooked capability of AIVs to handle multiple jobs simultaneously, which significantly improves both energy efficiency and job tardiness.

Our application of reinforcement learning, specifically the Multi-Agent Deep Q-Network (MADQN), demonstrates the method's suitability for real-time scheduling, offering a rapid response to dynamic changes. While it may not match the precision of exact optimization methods, it provides a favorable balance between speed and accuracy, particularly in large-scale problems.

However, this study has limitations. The uniqueness of the case study, with its complex constraints, meant there was no established benchmark, such as linear programming or metaheuristics, for direct comparison. We only tested one reinforcement learning method, DQN, though other methods like policy-based approaches could be worth exploring. Additionally, we employed a layer-based communication structure between agents, but other communication methods, such as Graph Neural Networks (GNN) or attention mechanisms, which are increasingly applied in the literature, could also be considered.

While we accounted for AIVs with a capacity of two jobs, higher-capacity AIVs could introduce further complexities, particularly in determining the most efficient policies for loading and unloading—whether to pick up all jobs first and then unload, or to handle them one by one. Furthermore, our study assumed a recharging policy where AIVs recharge when their battery level drops below 40\%. Exploring different charging strategies could provide insights into their effects on production objectives. Finally, while the proposed approach succeeded across different problem scales and scenarios with varying numbers of job arrivals, it was not tested under differing shop configurations. Shop configurations, defined by the number of job types, machines, and AIVs, could provide valuable proof of the method's ability to generalize to various circumstances.

\section{Conclusion}
This study presents a solution for scheduling autonomous internal logistic vehicles in smart manufacturing environments using multi-agent deep Q-network (MADQN) with a layer-based communication channel (LBCC). The proposed approach effectively addresses three key objectives: minimizing the total tardiness of jobs, reducing the number of tardy jobs, and decreasing the total energy consumption of vehicles. Our comprehensive analysis, comparing the method against nine well-known heuristics, highlights the superior performance and adaptability of the MADQN with LBCC.

The method’s compatibility with flexible job shops exhibiting dynamic behaviors, such as dynamic job arrivals and workstation unavailabilities, underscores its practical applicability. Furthermore, the proposed approach ensures consistent results across various job shop layouts and larger problem instances, making it an adaptable tool for real-world applications. For practical applications, it is recommended to integrate the proposed method into fleet management software. This integration allows the algorithm to offer decision recommendations each time a new order is received. These recommendations can be reviewed by experts to finalize vehicle and machine selection decisions, thereby improving both the efficiency and reliability of the decision-making process. 

Despite the promising results of this study, several avenues for future research are evident. Future work should explore a broader range of reinforcement learning methods, including policy-based approaches, to enhance the robustness of the proposed solution. Additionally, investigating alternative communication methods for multi-agent systems, such as attention mechanisms, could provide new insights and improve performance.The study focused on AIVs with a capacity of two jobs, but exploring higher-capacity AIVs could address new challenges in optimizing loading and unloading strategies. Different recharging policeies for AIVs should also be evaluated to assess their impact on production efficiency. Lastly, testing the proposed approach across various shop configurations, defined by different numbers of job types, machines, and AIVs, could validate its generalizability and effectiveness in diverse settings.

%\section{Limitations}
%\hl{
%In this work, a multi-agent deep reinforcement learning (MADRL) framework with a Layer-Based Communication Channel (LBCC) is proposed to solve the dynamic Flexible Job Shop Scheduling Problem (FJSSP). The experimentation primarily focuses on evaluating the performance of the Deep Q-Network (DQN) algorithm within this framework. However, the evaluation can be expanded to include other DRL algorithms such as Proximal Policy Optimization (PPO), Deep Deterministic Policy Gradient (DDPG), and others, allowing for a more comprehensive assessment of the approach.

%Moreover, the current benchmarks of the proposed approach have been executed within a fixed shop configuration, where the number of machines, job types, and AIVs remain constant. To ensure the generality and robustness of the proposed method, future evaluations can alter these parameters, testing the approach across various shop sizes and configurations. This would further validate the scalability and adaptability of the proposed MADRL framework in different dynamic environments.}
%\hl{\cite{HOSSEINI2024} hasn't been published yet}
\bibliographystyle{unsrt}
\bibliography{example} 
\end{document}